\documentclass[twocolumn,9pt]{article} 

\usepackage[square,numbers,sort&compress,comma]{natbib}

\usepackage{amsmath, bm}
\usepackage{amssymb}
\usepackage{caption}
\usepackage{graphicx}
\usepackage{latexsym}
\usepackage{times}
\usepackage[pagewise]{lineno}
\usepackage{hyperref}
\usepackage{glossaries}
\usepackage[margin=1cm]{geometry}
\usepackage{subcaption}
\usepackage{pgfplots}
\usepackage{tikz}
\usetikzlibrary{calc}
\usetikzlibrary{arrows,decorations.markings}
\usepackage{xcolor}
\usepackage{bm}
\usepackage{array}
\usepackage{tabularx}
\usepackage{makecell}
\usepackage{yhmath}

\usepackage{glossaries-extra}
\setabbreviationstyle[acronym]{long-short}
\glssetcategoryattribute{acronym}{nohyperfirst}{true}

\newacronym{ATF}{ATF}{Artificially Thickened Flame}
\newacronym{CFD}{CFD}{Computational Fluid Dynamics}
\newacronym{CFL}{CFL}{Courant-Friedrichs-Lewy}
\newacronym{DG}{DG}{Discontinuous Galerkin}
\newacronym{DNS}{DNS}{Direct Numerical Simulation}
\newacronym[\glslongpluralkey={Dispersion relation}]{DR}{DR}{Dispersionsrelation}

\newacronym{EBU}{EBU}{Eddy Break-Up}
\newacronym{EP}{EWP}{Eigenwert Problem}
\newacronym{FDF}{FDF}{Flame Describing Function}
\newacronym{FE}{FE}{Finite Elemente}
\newacronym{FEM}{FEM}{Finite Element Method}
\newacronym{FES}{EXP}{Chair of Fluid Dynamics}
\newacronym{FELiCS}{FELiCS}{Finite Element Linearized Combustion Solver}

\newacronym{FSD}{FSD}{Flame Surface Density}
\newacronym{FTF}{FTF}{Flame Transfer Function}
\newacronym{GEVP}{GEVP}{Generalized Eigenvalue Problem}
\newacronym{HPC}{HPC}{High Performance Computing}
\newacronym{IOA}{IOA}{Input-Output Analysis}
\newacronym{IDP}{IDP}{Inverse Design Procedure}
\newacronym{ISTA}{ISTA}{Institut of Fluid Dynamics and Technical Acoustics}
\newacronym{IWB}{IWB}{Institute for Machine Tools and Industrial Management}
\newacronym{KH}{KH}{Kelvin-Helmholtz}
\newacronym{LA}{LA}{Linear Analysis}
\newacronym{LW}{LW}{Lax-Wendroff}
\newacronym{LES}{LES}{Large Eddy Simulation}
\newacronym{LHS}{LHS}{left hand side}
\newacronym{LNSE}{LNSE}{Lineaerized Navier-Stokes Equations}
\newacronym{LMFA}{LMFA}{Linearized Mean Field Analysis}
\newacronym{LSA}{LSA}{Linear Stability Analysis}
\newacronym{MMC}{MMC}{Multiple Mapping Conditioning}
\newacronym{NOx}{NOx}{Nitrogen oxides}
\newacronym{NSCBC}{NSCBC}{Navier-Stokes Characteristic Boundary Conditions}
\newacronym{PDE}{PDE}{Partial Differential Equation}
\newacronym{PBF-LB/M}{PBF-LB/M}{powder bed fusion of metals using a laser beam}
\newacronym{PLIF}{PLIF}{planar laser-induced fluorescence}
\newacronym{PI}{PI}{Principle Investigator}
\newacronym{PIV}{PIV}{Particle Image Velocimetry}
\newacronym{PINN}{PINN}{Physics Informed Neural Network}
\newacronym{PVC}{PVC}{Precessing Vortex Core}
\newacronym{RA}{RA}{Resolvent Analysis}
\newacronym{RANS}{RANS}{Reynolds-averaged Navier Stokes}
\newacronym{RHS}{RHS}{right hand side}
\newacronym{RST}{RST}{Reynolds-Stress Transport}
\newacronym{RQ}{RQ}{Research Question}
\newacronym{SGS}{SGS}{Subgrid Scale}
\newacronym{SPL}{SPL}{Sound Pressure Level}
\newacronym{SPP}{SPP}{Priority Program}
\newacronym{SPOD}{SPOD}{Spectral Proper Orthogonal Decomposition}
\newacronym{eSPOD}{eSPOD}{extended Spectral Proper Orthogonal Decomposition}

\newacronym{TAI}{TAI}{Thermoacoustic Instabilities}
\newacronym{TD}{TD}{thermodiffusive}
\newacronym[\glslongpluralkey={Thermodiffusive Instabilitäten}]{TDI}{TDI}{Thermodiffusive Instabilität}
\newacronym{TF}{TF}{Transfer Function}
\newacronym{TFC}{TFC}{Turbulent Flame-speed Closure}
\newacronym{TKE}{TKE}{Turbulent-Kinetic-Energy}
\newacronym{TTGC}{TTGC}{Two-Step Taylor Galerkin Scheme C}
\newacronym{TUB}{TUB}{TU Berlin}
\newacronym{TUM}{TUM}{TU Munich}
\newacronym{UCI}{UCI}{University of California Irvine}
\newacronym{WA}{WA}{work area}
\newacronym{Wasserstoff}{$\text{H}_2$}{Wasserstoff}

\newacronym{WHI}{WHI}{Wiener-Hopf Inversion}
\newacronym{felics}{FELiCS}{Finite Element Linear Combustion Solver}
\newacronym{fenics}{FEniCS}{Finite Element Computational Software}
\newacronym{TFM}{TFM}{Thickened Flame Model}

\topmargin - 12pt 
\renewenvironment{abstract}%
              {
               \small
               {\bfseries \abstractname}
               \par
               \vspace{10pt}
              }

\renewcommand\abstractname{Abstract}

\newcommand{\nomenclature}
              [1]
              {
               \bgroup
               \flushleft
               \small\bf
               #1
               \par
               \egroup
              }

\renewcommand{\section}
              [1]
              {
               \bgroup
               \flushleft
               \small\bf
               \refstepcounter{section}
               \arabic{section}. #1
               \par
               \egroup
              }

\renewcommand{\subsection}
              [1]
              {
               \bgroup
               \flushleft
               \small\em
               \refstepcounter{subsection}
               \arabic{section}.
               \arabic{subsection}. #1
               \par
               \egroup
              }

\renewcommand{\subsubsection}
              [1]
              {
               \bgroup
               \flushleft
               \small\em
               \refstepcounter{subsubsection}
               \arabic{section}.
               \arabic{subsection}.
               \arabic{subsubsection}. #1
               \par
               \egroup
              }

  \newcommand{\acknowledgement}
              [1]
              {
               \bgroup
               \flushleft
               \small\bf
               #1
               \par
               \egroup
              }

  \newcommand{\sectionbib}
              [1]
              {
               \bgroup
               \flushleft
               \small\bf
               #1
               \par
               \egroup
              }

\newcommand{\bs}[1]{\boldsymbol{#1}}
\newcolumntype{C}[1]{>{\centering\arraybackslash}p{#1}}

\newcolumntype{C}[1]{>{\centering\arraybackslash}m{#1}}
\begin{document}



\small
\baselineskip 10pt

\setcounter{page}{1}
\title{\LARGE \bf Modeling of Reaction Dynamics in a Turbulent
Hydrogen--Air Slot Flame Using Resolvent Analysis}

\author{{\large Anant Rajeev Talasikar $^{a,*}$, Marina Matthaiou$^{a}$, Michael Gauding$^{b}$, Heinz Pitsch$^{b}$,}\\{\large Thomas Ludwig Kaiser$^{a}$} \\[10pt]
        {\footnotesize \em $^a$ Institute of Fluid Dynamics and Technical Acoustics (ISTA), TU Berlin}\\[-5pt]
        {\footnotesize \em $^b$Chair and Institute for Technical Combustion, RWTH Aachen}}

\date{}  

\twocolumn[\begin{@twocolumnfalse}
\maketitle
\rule{\textwidth}{0.5pt}
\vspace{-5pt}

\begin{abstract} 

This work applies Resolvent Analysis (RA) to study the dynamics of a hydrogen–air slot flame with a Reynolds number of 5500, a Karlovitz number of 20, and an equivalence ratio of 0.4. Direct Numerical Simulations (DNS) data are analyzed using shifted Spectral Proper Orthogonal Decomposition (SPOD), and the resulting structures are compared with optimal resolvent responses obtained from the linearization of a RANS-EBU reaction rate model. Both SPOD and RA show that the flow dynamics are dominated by Kelvin–Helmholtz wave packets over a broad frequency range, particularly between 300 and 1000 Hz. This behavior is reflected in the resolvent gains and SPOD eigenvalues, which exhibit consistent amplification within this range. The velocity fluctuation mode shapes predicted by RA agree well with the SPOD modes. However, the corresponding mode shapes for the progress variable and heat release show weaker agreement. To address this limitation, the study introduces a generalized active-flame closure calibrated with high-fidelity data, which remains compatible with the linearized framework and improves the agreement with SPOD modes. Overall, the results indicate that thermodiffusive instabilities in turbulent hydrogen flames do not hinder the applicability of the active-flame resolvent approach.

\end{abstract}
\glsresetall
\vspace{10pt}

{\bf Novelty and significance statement}

\vspace{10pt}

This study demonstrates for the first time that Resolvent Analysis (RA) can be successfully applied to a turbulent hydrogen–air slot jet flame. To improve the modeling of the interaction between the turbulence and the flame, we introduce an active-flame resolvent framework based on a statistically derived algebraic flame closure model. The study demonstrates that this model performs better in reproducing the dynamics of the turbulent hydrogen--air slot flame than previously investigated models~\cite{kaiser2023modelling}. Finally, the results indicate that, even in hydrogen flames affected by thermodiffusive effects, the linear resolvent framework remains predictive regarding the dominant dynamics in the flame. Therefore, this study opens new pathways for applying active-flame RA to different turbulent reacting flows, based on the enhancement in the predictive capability and adaptability of data-informed algebraic closures.

\vspace{5pt}
\parbox{1.0\textwidth}{\footnotesize {\em Keywords:} Resolvent Analysis; Turbulent Flame; Spectral Proper Orthogonal Decomposition; Direct Numerical Simulation;}
\rule{\textwidth}{0.5pt}
*Corresponding author.
\vspace{5pt}
\end{@twocolumnfalse}] 

\section{Introduction\label{sec:introduction}} \addvspace{5pt}



Linear mean-field analysis, which studies the governing equations of the flow linearized around their temporal mean solution, has advanced the understanding of various flow configurations. \gls{RA}, a form of linear mean-field analysis, identifies dominant amplification mechanisms, particularly in turbulent flows. Its origins are in studies of turbulent transition \cite{trefethen1993hydrodynamic, butler1992three, farrell1993stochastic, reddy1993energy, reddy1993pseudospectra}. \gls{RA} has since contributed to a better understanding of turbulence in boundary layers \cite{cossu2009optimal, sipp2013characterization}, jets \cite{garnaud2013preferred, beneddine2017unsteady, schmidt2018spectral}, airfoil wakes \cite{abreu2018wavepackets, demange_resolvent_2024}, channels \cite{abreu2020resolvent, morra2019relevance}, pipes, and Couette flows \cite{abreu2020spectral, mckeon2010critical, hwang2010amplification}, as well as high-Mach number jet studies on flame noise \cite{pickering2020resolvent, pickering2021resolvent}.
\addvspace{-2pt}


While linear analysis has long been used for nonreacting flows, its application to flame configurations is relatively recent. Most studies of laminar premixed flames focus on thermoacoustics, such as Blanchard et al.~\cite{blanchard_response_2015}, who used \gls{RA} to show that flame wrinkling generates hydrodynamic disturbances that propagate through the reactant flow and dominate the low-frequency response of methane–air flames. Wang et al.~\cite{wang2022linear} applied a similar approach to study intrinsic instabilities in laminar flames. Avdonin et al.~\cite{avdonin2019thermoacoustic} developed linearized reactive flow formulations that capture flame flow interactions and predict \glspl{FTF} and thermoacoustic eigenmodes. In contrast to global \glspl{FTF}, Meindl et al.~\cite{meindl2021spurious} showed that local \glspl{FTF} avoid spurious entropy perturbations and the monolithic linearized reactive flow approaches correctly capture flame-front motion and unsteady heat release dynamics.


Similarly, linearized analyses are increasingly used to study turbulent combustion dynamics and coherent structures. In one methodology, the passive flame approach, the flame is considered in the linear framework by incorporating density variations across the flame front but neglecting fluctuations in heat release. The work of Manoharan et al.~\cite{manoharan2015absolute} used this approach to identify vorticity fluctuations and baroclinic torque as drivers of convective instabilities in turbulent shear layers. Also, Kaiser et al. (2019)~\cite{kaiser_prediction_2019} used a passive flame \gls{RA} to predict the hydrodynamic response of a swirled jet flame to acoustic forcing, while Casel et al.~\cite{casel2022resolvent} showed that \gls{RA} with a passive flame approach reproduces dominant Kelvin–Helmholtz structures in turbulent flames. Kaiser et al. (2023)~\cite{kaiser2023modelling} introduced an active flame approach, where the fluctuations in heat release are included in the linear model. In this context, they demonstrated that linearization of \gls{DNS} chemistry in a round methane-air jet flame leads to closure issues, similar to those in \gls{RANS}. Instead, linearizing \gls{RANS} reaction models provides a viable alternative, accurately capturing flame dynamics under acoustic actuation. Recently, Chauhan et al.~\cite{chauhan_modeling_2025, chauhan:hal-05499375} applied both passive and active flame approaches to accurately predict \gls{FTF} gain and phase in turbulent premixed jet flames.

All previous linear studies of flame configurations have focused on round methane–air jets. It is currently unknown if an application of the existing approach to hydrogen flames is possible since thermodiffusive instabilities~\cite{berger2025combustion}, which significantly impact flame dynamics in lean hydrogen flames, are not explicitly modeled in the current approach. If direct application is not feasible, the method must be modified accordingly. Furthermore, all previous linear analyses on reacting flows have focused on rationally symmetric configurations. To address this research gap, this study analyzes a turbulent lean hydrogen–air slot flame, using \gls{SPOD} and active-flame \gls{RA}, with emphasis on the adaptability of \gls{RA} when informed by high-fidelity data. To limit the scope of the study, the analysis is restricted to zero wavenumber structures in the homogeneous direction of the configuration.

This paper is organized as follows. Sections~\ref{sec:dns} and~\ref{ch:theory:spod} outline \gls{DNS} and introduce \gls{SPOD}, respectively. Section~\ref{sec:BGandTH} introduces the active-flame closure in the \gls{RA} framework and its calibration to \gls{DNS} data. Section~\ref{sec:Results and discussion} presents the results, starting with \gls{SPOD}. Then the linear reaction model is tested in an a priori analysis, before the \gls{RA} results are presented, compared against the \gls{SPOD} results and interpreted. Section~\ref{conclusion} summarizes the study and highlights its scope.




\section{Direct Numerical Simulations \label{sec:dns}} \addvspace{10pt}
\begin{figure}
    \centering
    \includegraphics[width=0.9\linewidth]{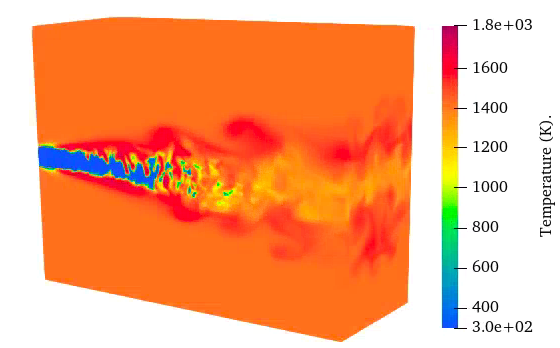}
    \caption{Temperature field from the \gls{DNS} data.}
    \label{fig:DNS-TEMP}
\end{figure}

A \gls{DNS} of a lean premixed turbulent hydrogen--air flame in a piloted slot burner configuration has been performed using the CIAO code \cite{desjardins2008high}, which solves the reacting Navier--Stokes equations in the low-Mach number limit. The numerical domain is shown in Fig.~\ref{fig:DNS-TEMP}, where the colormap indicates the instantaneous temperature. A mixture-averaged model is employed to determine the thermal conductivity. The viscosity and molecular diffusion coefficients are computed assuming constant Lewis numbers, which are evaluated in the burned gas. The slot flame features an equivalence ratio of $\phi=0.4$, an unburnt temperature of $T_u=298~K$, and a pressure of 1~atm. The jet Reynolds number equals 5,500, based on the slot width of $\text{H}=4$~mm and the bulk velocity of $U=24$~m/s. Further details on the setup and flame configuration are provided by Berger et al.~\cite{berger2024effects}.

A key challenge in this study is obtaining sufficiently long \gls{DNS} time series with high spatial and temporal resolution for \gls{SPOD}. The dataset used here consists of 300 three-dimensional snapshots sampled at intervals of $5\times10^{-5}~\text{s}$.


\section{Spectral Proper Orthogonal Decomposition\label{ch:theory:spod}} \addvspace{10pt}
In this work, coherent structures are extracted using \gls{SPOD}~\cite{Blanco_Martini_Sasaki_Cavalieri_2022}. Prior to the decomposition, the temporal mean is subtracted, the fluctuations are split into symmetric and anti-symmetric components with respect to the slot symmetry plane (x--y plane), and averaged in the homogeneous z--direction to enforce zero wave number in z--direction. Then, a space-time shift is performed \cite{Blanco_Martini_Sasaki_Cavalieri_2022}, accounting for the convection of the coherent structures. The resulting snapshots are segmented into $N_\text{seg}$ blocks with overlap $N_\text{over}$, Fourier transformed, and used to form a Welch-type estimate of the cross-spectral density tensor. Its eigen-decomposition yields \gls{SPOD} eigenvalues and eigenvectors. The eigenvectors describe the spatial shape of a coherent structure, while the eigenvalue gives an indication of its respective energy content. A clearly separated leading eigenvalue indicates low-rank behavior and a dominant wave packet. There is a strong conceptual link between \gls{RA} and \gls{SPOD}, allowing flow configurations to validate \gls{RA} gains against \gls{SPOD} eigenvalues and the \gls{RA} responses against the \gls{SPOD} eigenvectors \cite{towne_spectral_2018}. A detailed introduction to \gls{SPOD} is given in Schmidt and Colonius~\cite{schmidt2020guide}.

\section{Linear Modeling Theory\label{sec:BGandTH}} \addvspace{10pt}
\subsection{Linearization of the Governing Equations\label{subsec:Linearization}} \addvspace{5pt}

The nonlinear governing equations, which will be linearized in the remainder of this section, are the Navier--Stokes equations and a transport equation for the reaction progress variable, $c$. These are given by
\begin{subequations} \label{eq:NL_Eqs}
    \begin{align}
        \begin{split}
            &\rho \biggl(\frac{\partial \bs{u}}{\partial t} + (\bs{u}\cdot\nabla)\bs{u}\biggr) = -\nabla p + \nabla \cdot \mu \biggl(\nabla \bs{u} \biggr. \\ &\biggl. + (\nabla \bs{u})^T + -\frac{2}{3} (\nabla \cdot \bs{u}) I \biggr),  
        \end{split} \label{NL_MOMENTUM}
        \\
        &\frac{\partial \rho}{\partial t} + \nabla \cdot (\rho \bs{u}) = 0,
        \\
        &\rho\frac{\partial c}{\partial t} + \rho \bs{u} \cdot \nabla c = \nabla \cdot \Gamma \nabla c + \dot{\Omega}. \label{NL_Progress}
    \end{align}
\end{subequations}

\noindent Here, $\mu$ is the dynamic viscosity and $\Gamma$ is the diffusivity for the progress variable. A low-Mach number assumption allows for variable density while the flow remains incompressible. The assumptions of equal and constant thermophysical properties for all relevant chemical species and the adiabatic numerical setup of the \gls{DNS} directly link the reaction progress variable to the temperature, $T$, via a linear relation and to the density, $\rho$, based on the perfect gas law

\begin{align}
    \rho = \frac{p_0}{RT}, \quad T = T_u + (T_b - T_u)c. \label{eq:fromCToRho}
\end{align}
The thermodynamic pressure, $p_0$, and the specific gas constant, $R$, are assumed to be constant, while $T_\text{u} = 298 \, \text{K}$ and $T_\text{b} = 1418 \, \text{K}$ are the temperature of the unburnt mixture and the adiabatic flame temperature, respectively. The reaction modeling will be discussed in Section~\ref{subsubsec:reactionModelling}.
To linearize the set of Eqs.~(\ref{eq:NL_Eqs})-(\ref{eq:fromCToRho}), a triple decomposition as in \cite{hussain1970mechanics} and \cite{reynolds1972mechanics} is applied to the state variable $q = [\bs{u}, p, \rho, T, c]$ as 
\begin{equation}
    q = \overline{q} + \widetilde{q} + q', \label{eq:TripDecomp}
\end{equation} 
where $(\overline{\cdot})$ denotes the temporal mean, $(\widetilde{\cdot})$ denotes the coherent fluctuation, and $(\cdot')$ denotes the stochastic fluctuation. Coherent fluctuations of the state variables are obtained by the difference of the phase average and the time average given by
\begin{equation}
    \widetilde{q} = \langle q \rangle - \overline{q}. \label{eq:phaseEqs}
\end{equation} 

Substituting Eqs.~(\ref{eq:TripDecomp}),~(\ref{eq:phaseEqs}) into Eqs.~(\ref{eq:NL_Eqs})–(\ref{eq:fromCToRho}) and neglecting nonlinear terms in the coherent fluctuations yields a linearized system for the coherent perturbations. The resulting equations contain unclosed terms such as $\widetilde{\rho(\bs{u}'\cdot \nabla)\bs{u}'}$ and $\widetilde{\rho \bs{u}' \cdot \nabla c'}$, which represent fluctuations of Reynolds stresses and turbulent transport in $\widetilde{c}$ during the coherent fluctuation $\widetilde{q}$ and are of the same order as the linear perturbations \cite{reynolds1972mechanics}. In this study, these terms are modeled through turbulent viscosity $\mu_t$ and turbulent diffusivity $D_t$, following \cite{viola2014prediction} and \cite{kaiser2021modeling}. Finally, applying the ansatz 
\begin{equation}
    \widetilde{q} = \widehat{q}~\text{exp}(-i\omega t) + \text{c.c.}, \label{eq:ansatz}
\end{equation}
where, in this study, the wavenumber in the homogeneous z--direction is zero. In Eq.~\ref{eq:ansatz}, $\omega$ is the real angular frequency that transforms the equations into the frequency domain, yielding

\begin{subequations} \label{eq:linearized_equations}   
\begin{align}
    \begin{split}
        &-i \omega \bar{\rho} \bs{\hat{u}} + \bar{\rho} ((\hat{\bs{u}} \cdot \nabla)\bar{\bs{u}} + (\bar{\bs{u}} \cdot \nabla )\hat{\bs{u}} ) \\ &+ \hat{\rho} (\bar{\bs{u}} \cdot \nabla) \bar{\bs{u}}  = - \nabla \hat{p}  + \nabla \cdot (\mu + \mu_t ) \biggl(\nabla \hat{\bs{u}} \biggr. \\ &\left. + (\nabla \hat{\bs{u}})^T - \frac{2}{3} (\nabla \cdot \hat{\bs{u}}) \textbf{I} \right) + \widehat{\mathbf{f}}_u,
    \end{split} \label{linearized_momentum}
    \\
    &-i \omega \hat{\rho} + \nabla \cdot(\hat{\rho} \bar{\bs{u}}) + \nabla \cdot (\bar{\rho} \hat{\bs{u}}) = 0,
    \\
    \begin{split}
    & -i \omega \bar{\rho} \hat{c} + \hat{\rho} \bar{\bs{u}} \cdot \nabla \bar{c} + \bar{\rho}\hat{\bs{u}} \cdot \nabla \bar{c} + \bar{\rho}\bar{\bs{u}} \cdot \nabla \tilde{c} = \\ & \nabla \cdot \bar{\rho}(D + D_t) \nabla \hat{c} + \hat{\dot{\Omega}},
    \end{split} \label{linearized_progress}
    \\
    &\hat{\rho} = - \frac{\bar{\rho}}{\bar{T}} \hat{T}, \hspace{0.5cm} \hat{T} = (T_b - T_u) \hat{c}. \label{linearized_eos}
\end{align}
\end{subequations}
This set of equations governs the fluctuations in state variables ($\widehat{\bs{u}}$, $\widehat{p}$, and $\widehat{c}$), while the temporal mean states ($\overline{\bs{u}}$, $\overline{c}$, $\overline{T}$, and $\overline{\rho}$) are obtained from the \gls{DNS}. 
The mean progress variable is defined as 
\begin{align}
    \overline{c} = \frac{Y_{H_2 ,u}-\overline{Y}_{H_2}}{Y_{H_2, u} - Y_{H_2 ,b}}, \label{progress}
\end{align}
 
\noindent where $\overline{Y}_{H_2}$ is the temporally averaged mass fraction of hydrogen, while $Y_{H_2,u}$ and $Y_{H_2,b}$ are the mass fractions of hydrogen in the unburnt mixture and at chemical equilibrium, respectively. The volume force term $\widehat{\mathbf{f}}_u$ is attributed to the triadic interaction with coherent structures at other frequencies. Finally, a model for the source term $\hat{\dot{\Omega}}$ in the linearized transport equation of the progress variable is needed. This will be discussed in detail in Section~\ref{subsubsec:reactionModelling}.

\subsection{Resolvent Analysis} \addvspace{5pt}
The objective of \gls{RA} is to find the pattern in a volume forcing, here $\widehat{\mathbf{f}}$, which produces the strongest response in the state variables governed by the set of Eqs.~(\ref{eq:linearized_equations}). To obtain this forcing, we rearrange the set of linearized Eqs.~(\ref{eq:linearized_equations}) in symbolic notation as
\begin{align}
    \omega B \hat{q}  + L\hat{q} = \hat{f} \implies \hat{q} = \underbrace{(L-\omega B)^{-1}}_{R(\omega)}  \hat{f}, \label{eq:EVP}
\end{align}
where $R(\omega)$ is the resolvent operator, which projects an arbitrary forcing on the linear response $\widehat{q}$ of the flow. 

At the core of \gls{RA} is the definition of a gain function, $\sigma^2$, defined as
\begin{equation}
    \sigma^2 = \frac{\lVert\widehat{q}\rVert_{\widehat{q}}}{\lVert\widehat{f}\rVert_{\widehat{f}}},   \label{eq:gain}
\end{equation}
In general, the response and forcing norms, $\lVert\cdot\rVert_{\widehat{q}}$ and $\lVert\cdot\rVert_{\widehat{f}}$, are arbitrary and can be chosen, depending on the quantities of interest. In this study, they are chosen as the $\text{L}_2$ norm of the velocity components. By combining Eqs.~(\ref{eq:EVP}) and (\ref{eq:gain}), a singular value problem can be derived (see e.g., Beneddine et al.~\cite{beneddine2017unsteady}), which yields $\sigma_i$ as singular values, the optimal forcings, $\widehat{\Phi}_i$, as right singular vectors, and the optimal responses, $\widehat{\Psi}_i$, which correspond to the left singular vectors. Sorting the optimal forcings $\widehat{\Phi}_i$ in descending order according to their gains $\sigma_i$ enables identification of the spatial forcing structures that produce the strongest dynamical response in the flow regarding the norms defining the gain function in Eq.~(\ref{eq:gain}). When the leading gain is considerably larger than the remaining gains, the system is considered to exhibit low-rank behavior at that particular frequency. Under such conditions, the flow dynamics are primarily governed by the amplification mechanism associated with the dominant resolvent mode, characterized by the leading triplet consisting of the gain, optimal forcing, and response, i.e., ($\sigma_1$, $\Phi_1$, and $\Psi_1$).

\subsection{Flame Modeling in the Linear Framework \label{subsubsec:reactionModelling}} \addvspace{5pt}
\begin{figure}[t]
\centering
\begin{subfigure}{\linewidth}
  \centering
\includegraphics[width=\linewidth]{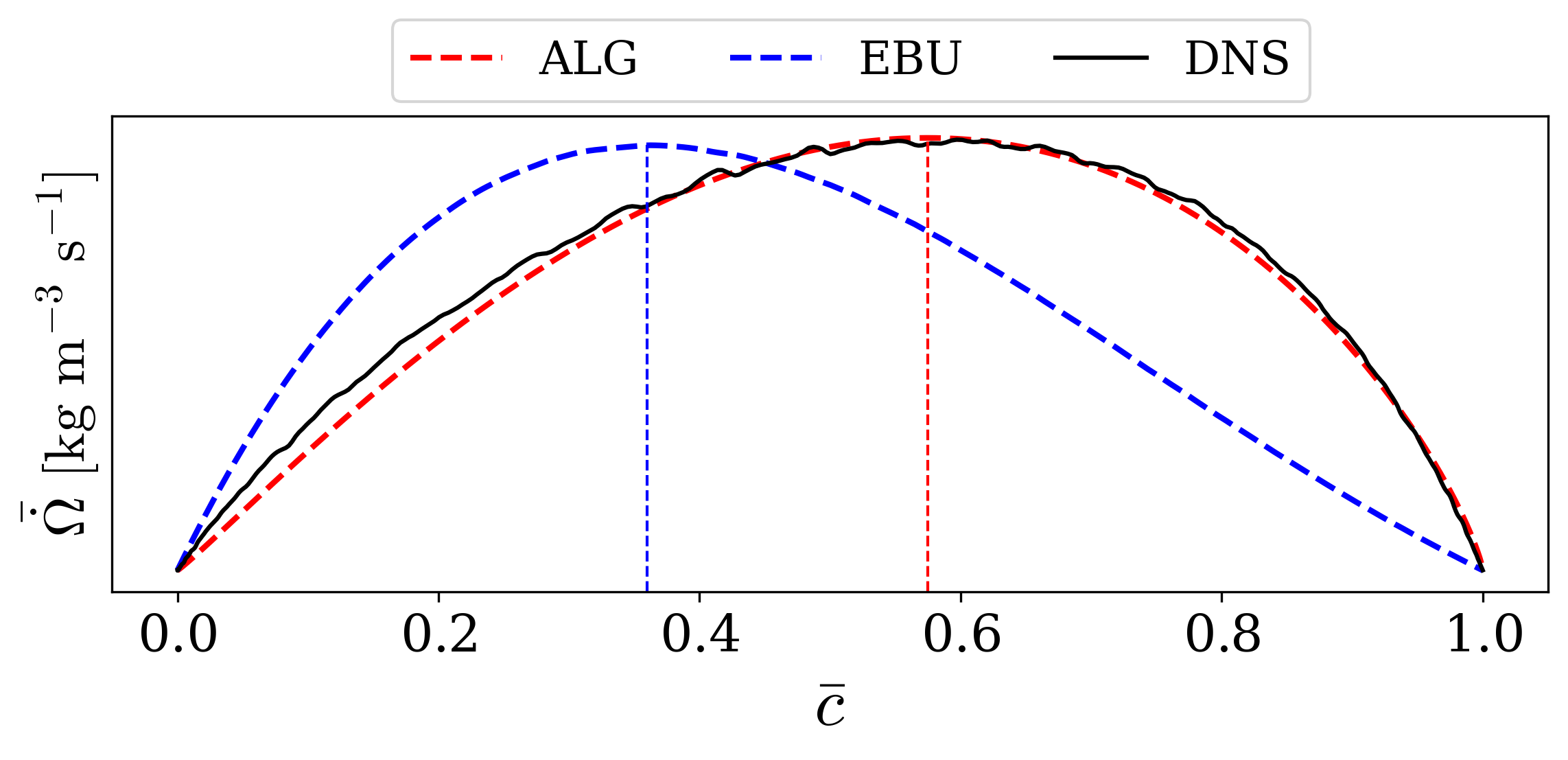}
\end{subfigure}
\caption{\footnotesize Nonlinear reaction rates of the EBU model and the algebraic model.}
 \label{fig:HRMs}
\end{figure}
Kaiser et al.~\cite{kaiser2023modelling} showed that directly linearizing \gls{DNS} or \gls{LES} combustion models leads to closure issues. In the cited work, this difficulty was avoided by instead linearizing a \gls{RANS}-\gls{EBU} model. Furthermore, the study demonstrated that, for a model to be suitable for linearization, it must accurately reproduce the temporal mean state of the flame. In their work, the \gls{EBU} model was shown to satisfy this requirement and is expressed as follows:
\begin{equation}
    \overline{\dot{\Omega}}_{\text{EBU}} = A_{\text{EBU}} \overline{\rho}\, \overline{c} (1-\overline{c}).\label{eq:EBU}
\end{equation}
Here, $A_{\text{EBU}}$ is an a priori unknown coefficient, which remains to be calibrated in the following. It is important to note that, in the present model, the temporal mean density $\overline{\rho}$ depends on the mean progress variable $\overline{c}$ (see Eq.~(\ref{eq:fromCToRho})), such that the reaction rate $\overline{\dot{\Omega}}_{\text{EBU}}$ becomes a function of the mean progress variable alone. However, an equivalent \gls{DNS} reaction rate is still required, since the \gls{DNS} formulation does not directly solve a transport equation for the progress variable. Therefore, using Eq.~\ref{progress}, the nonlinear source term in the progress variable transport equation, $\dot{\Omega}$, can be related to the heat release obtained from the \gls{DNS}, yielding an equivalent \gls{DNS} reaction rate as follows: 
\begin{equation}
    \overline{\dot{\Omega}}_{\text{DNS}}  = \frac{\text{HR}_{\text{DNS}}}{\text{LHV}(Y_{{H_2},u}-Y_{{H_2},b})}, \label{eq:tuning}
\end{equation}
where $\overline{\dot{\Omega}}_{\text{DNS}}$ is the mean equivalent \gls{DNS} reaction rate, LHV denotes the lower heating value of hydrogen, and $\text{HR}_{\text{DNS}}$ is the temporally averaged \gls{DNS} heat release. The variation of $\overline{\dot{\Omega}}_{\text{DNS}}$ along the symmetry plane is shown in Fig.~\ref{fig:HRMs} as a solid black line. The nonlinear \gls{EBU} model in Eq.~(\ref{eq:tuning}) is then calibrated by tuning the coefficient of $A_{\text{EBU}}$ such that the maximum value of the model aligns with the corresponding maximum value observed in the \gls{DNS}. The resulting relation $\overline{\dot{\Omega}}_{\text{EBU}}(\overline{c})$ for $A_{\text{EBU}} = 4470.72~\text{s}^{-1}$ is plotted against the \gls{DNS} reference in Fig.~\ref{fig:HRMs}. The model does not reproduce the \gls{DNS} trend well. In particular, $\overline{\dot{\Omega}}_{\text{DNS}}$ peaks at $\overline{c} \approx 0.57$, whereas the \gls{EBU} model peaks at $\overline{c} \approx 0.36$ (dashed vertical blue line in Fig.~\ref{fig:HRMs}). Following the argument of Kaiser et al.~\cite{kaiser2023modelling}, this discrepancy suggests that the \gls{EBU} model is unlikely to perform well in the linearized framework for the present configuration.

To improve the modeling of the reaction rate, this study proposes an alternative approach. According to Kaiser et al.~\cite{kaiser2023modelling}, a model suitable for linearization must reproduce the temporal mean state. This requirement, however, can also be satisfied by an algebraic expression calibrated to the temporal mean state obtained from high-fidelity data, which in the present case is the \gls{DNS}. In this study, the following expression will be used for this purpose.

\begin{equation}
    \overline{\dot{\Omega}}_{\text{ALG}} = A_{\text{ALG}} \overline{c}^{\alpha_1} (1-\overline{c})^{\alpha_2}. \label{eq:algebraic}
\end{equation}
A least-squares fit to the \gls{DNS} data yields $\alpha_1 = 1.08$ and $\alpha_2 = 0.8$. The nonlinear model is then calibrated to the peak of $\overline{\dot{\Omega}}{_\text{DNS}}$, yielding $A_{\text{ALG}}=2603.33$~kg~m\textsuperscript{-3}~s\textsuperscript{-1}. The resulting reaction rate, shown as a dashed red curve in Fig.~\ref{fig:HRMs}, closely matches the \gls{DNS} data. This approach leverages the physics inherent in the \gls{DNS} dataset and adds no extra numerical cost, since the temporal mean is already needed for \gls{RA}. However, it remains to be verified whether the linearized model retains predictive capability within the linear framework.

\noindent Finally, to close the set of linearized equations, the fluctuation in reaction rate $\widehat{\dot{\Omega}}$ is determined through the approximation of the nonlinear reaction rate, $\dot{\Omega}$ by a first-order Taylor series expansion around the temporal mean state. This yields the linearized reaction rate $\widehat{\dot{\Omega}} \approx \frac{\partial f(\overline{\Phi})}{\partial \overline{\Phi}} \widehat{\Phi}$, where $f(\overline{\Phi})$ represents either the \gls{EBU} or the algebraic model, and $\frac{\partial f(\overline{\Phi})}{\partial \overline{\Phi}}$ is the slope of the curves of either of these models in Fig.~\ref{fig:HRMs}. The linearized reaction models read
\begin{align}
    &\widehat{\dot{\Omega}}_{\text{EBU}} = A_{\text{EBU}}( \widehat{\rho}\, \overline{c} (1-\overline{c}) + \overline{\rho}\,  (\widehat{c}-2\overline{c}\, \widehat{c})), \label{eq:linEBU} 
    \\
    \begin{split}
    &\widehat{\dot{\Omega}}_{\text{ALG}} = A_{\text{ALG}} \bigl( \alpha_{1} \overline{c}^{(\alpha_{1} -1)} (1-\overline{c})^{\alpha_2} \bigr. \\ \bigl. &-  \alpha_2\overline{c}^{\alpha_1} (1-\overline{c})^{(\alpha_2 -1)})\widehat{c}. \label{eq:linAlg}
    \end{split}
\end{align}


\subsection{Numerical Approach of the Resolvent Analysis} \addvspace{5pt}
\begin{figure}
    \centering
    \includegraphics[width=0.9\linewidth]{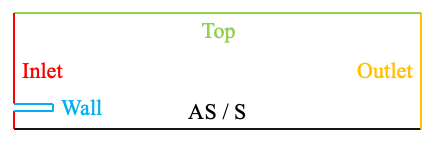}
    \caption{Boundaries of the 2D domain considered for \gls{RA}.}
    \label{fig:Domain}
\end{figure}

The \gls{RA} is performed using the \gls{FELiCS} software~\cite{kaiser_felics_2023}, which is based on the finite element package FEniCS~\cite{baratta_dolfinx_nodate}. The linearized system of Eqs.~(\ref{eq:linearized_equations}) is solved in 2D via discretization with Taylor–Hood finite elements, employing quadratic elements for $\widehat{\bs{u}}$ and linear elements for both $\widehat{p}$ and $\widehat{c}$. Spatial discretization is performed on an unstructured triangular mesh that partitions the computational domain and defines the associated mixed finite element function space. The trial and test functions are chosen from the same mixed finite element space, resulting in a continuous Galerkin formulation. This results in a \gls{GEVP}, which is solved in FELiCS using the linear solver \verb|KSPSolve()| of the \verb|PETSc| family \cite{petsc-efficient}. The boundary conditions for the 2D domain considered for \gls{RA} are listed in Table~\ref{tab:BCS}.



\begin{table}[ht!] \footnotesize
\centering

\begin{tabular}{|l|@{\hspace{2pt}}c@{\hspace{2pt}}|@{\hspace{2pt}}c@{\hspace{2pt}}|@{\hspace{2pt}}c@{\hspace{2pt}}|@{\hspace{2pt}}c@{\hspace{2pt}}|}
\hline
 & $\widehat{u}_x$ & $\widehat{u}_y$ & $\widehat{p}$ & $\widehat{c}$ \\ \hline

\textbf{\scriptsize Inlet} & $\widehat{u}_x = 0$ & $\widehat{u}_y = 0$ & $\widehat{p} = 0$ & $\widehat{c} = 0$ \\ \hline

\textbf{\scriptsize Outlet} & $\widehat{u}_x = 0$ & $\widehat{u}_y = 0$ & $\widehat{p} = 0$ & $\widehat{c} = 0$ \\ \hline

\textbf{\scriptsize Top} & $\widehat{u}_x = 0$ & $\widehat{u}_y = 0$ & $\widehat{p} = 0$ & $\widehat{c} = 0$ \\ \hline

\textbf{\scriptsize \makecell[l]{Anti-Symmetry\\ (AS)}} & $\widehat{u}_x = 0$ & $\frac{\partial \widehat{u}_y}{\partial \eta}=0$ & $\widehat{p} = 0$ & $\widehat{c} = 0$ \\ \hline

\textbf{\makecell[l]{\scriptsize Symmetry\\ (S)}} & $\frac{\partial \widehat{u}_x}{\partial \eta}=0$ & $\widehat{u}_y = 0$ & $\frac{\partial \widehat{p}}{\partial \eta}=0$ & $\frac{\partial \widehat{c}}{\partial \eta}=0$ \\ \hline

\textbf{\scriptsize Wall} & $\widehat{u}_x = 0$ & $\widehat{u}_y = 0$ & $\frac{\partial \widehat{p}}{\partial \eta}=0$ & $\widehat{c} = 0$ \\ \hline

\end{tabular}

\caption{Boundary conditions.}
\label{tab:BCS}

\end{table}


\section{Results and Discussion\label{sec:Results and discussion}} \addvspace{10pt}
\subsection{Data analysis results} \addvspace{10pt}

\begin{figure}[t]
\centering
\begin{subfigure}{\linewidth}
  \centering
\includegraphics[width=\linewidth]{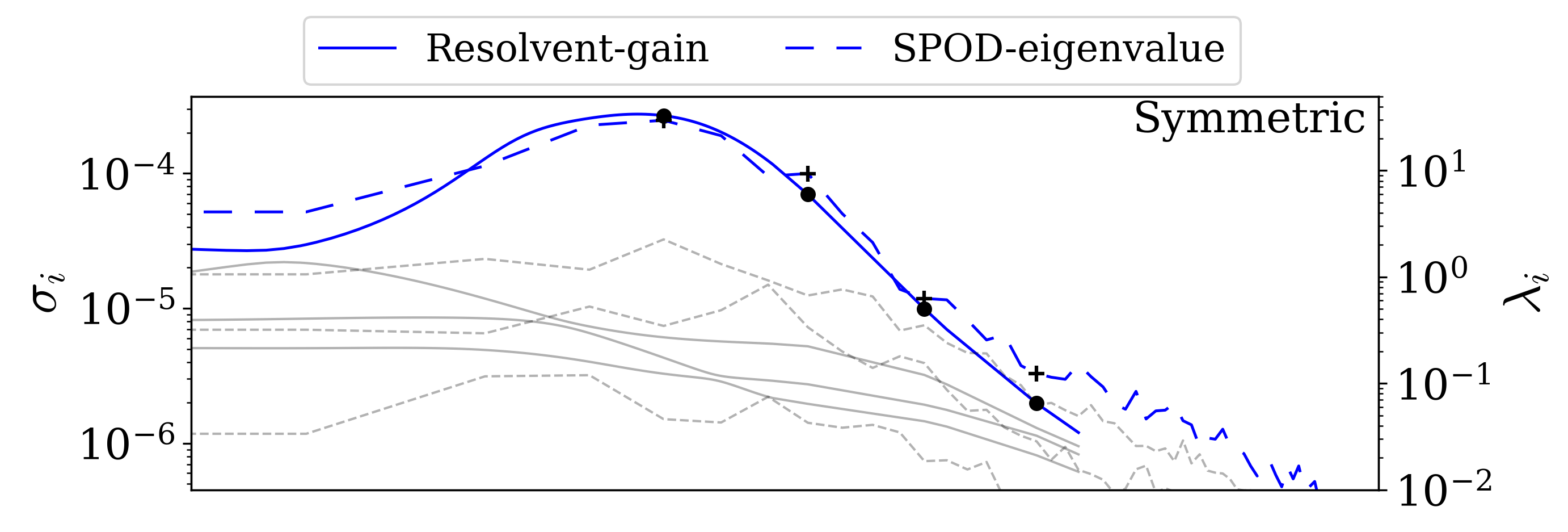}
  \label{fig:SPOD_sv_vel_s}
  \hfill
\begin{subfigure}{\linewidth}
  \centering
\includegraphics[width=\linewidth]{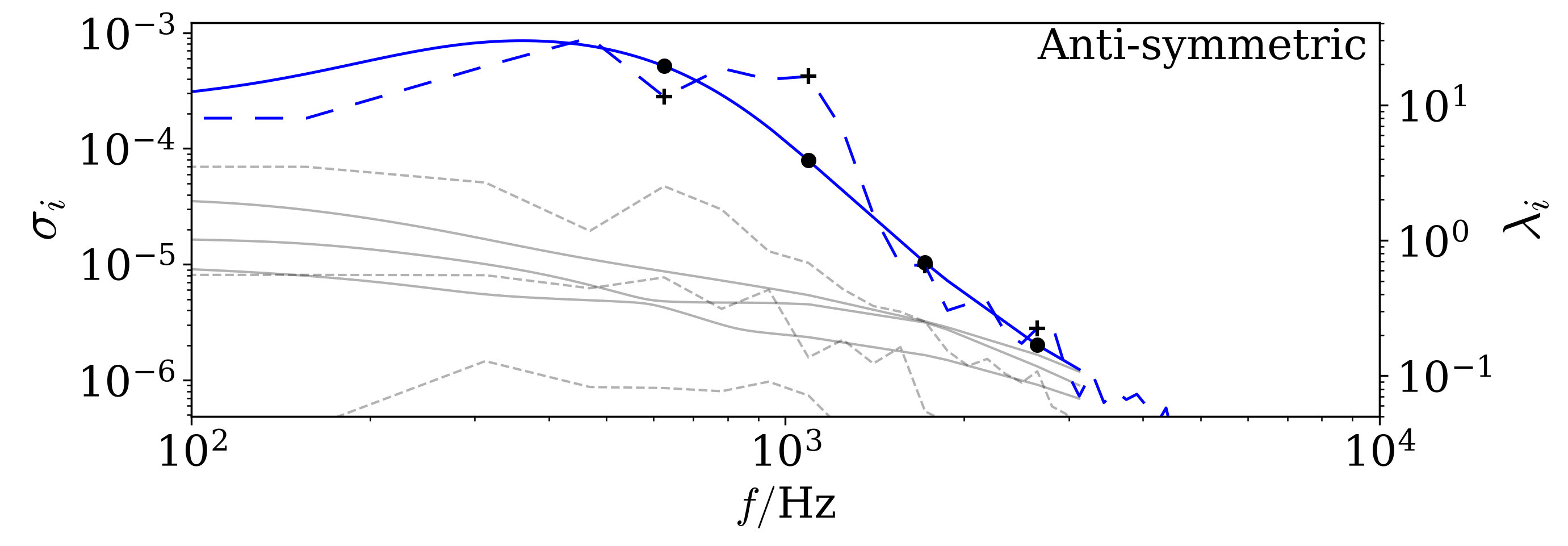}
  \label{fig:SPOD_sv_vel_as}
\end{subfigure}
\end{subfigure}
\caption{\footnotesize \gls{SPOD} eigenvalues for velocity fluctuations.}
\label{fig:gains_velocity}
\end{figure}

\begin{figure}[t]
\centering
\begin{subfigure}{\linewidth}
  \centering
  \hspace{-0.25cm} \includegraphics[width=0.975\linewidth]{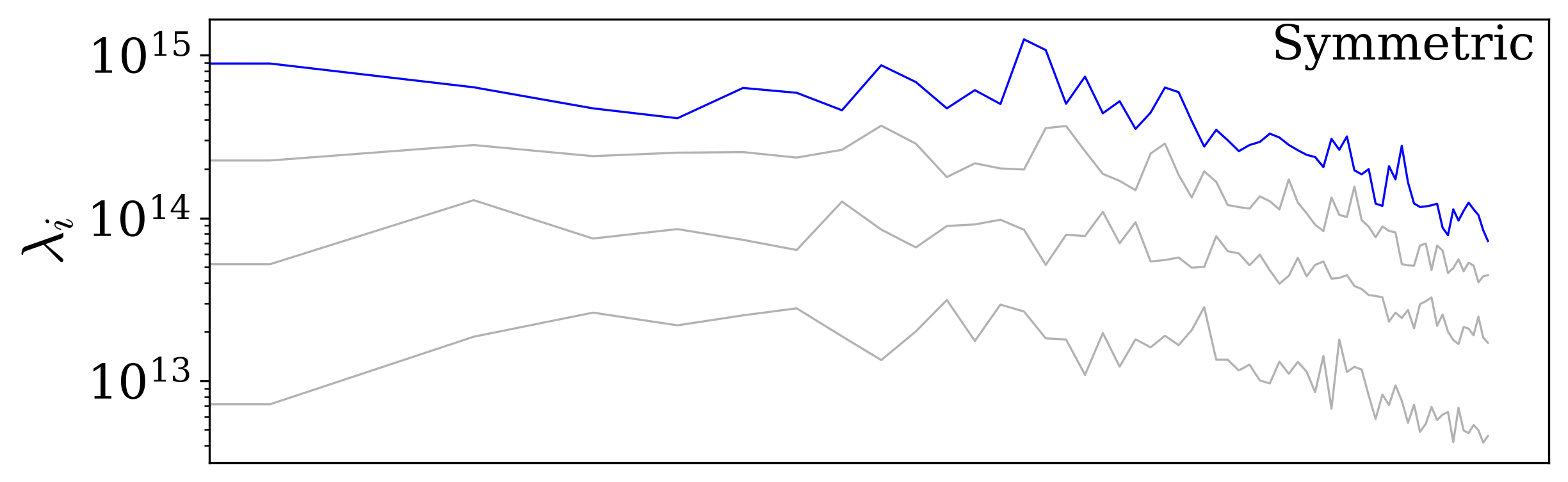}
  \label{fig:SPOD_sv_hr_s}
  \hfill
\begin{subfigure}{\linewidth}
  \centering
  \includegraphics[width=\linewidth]{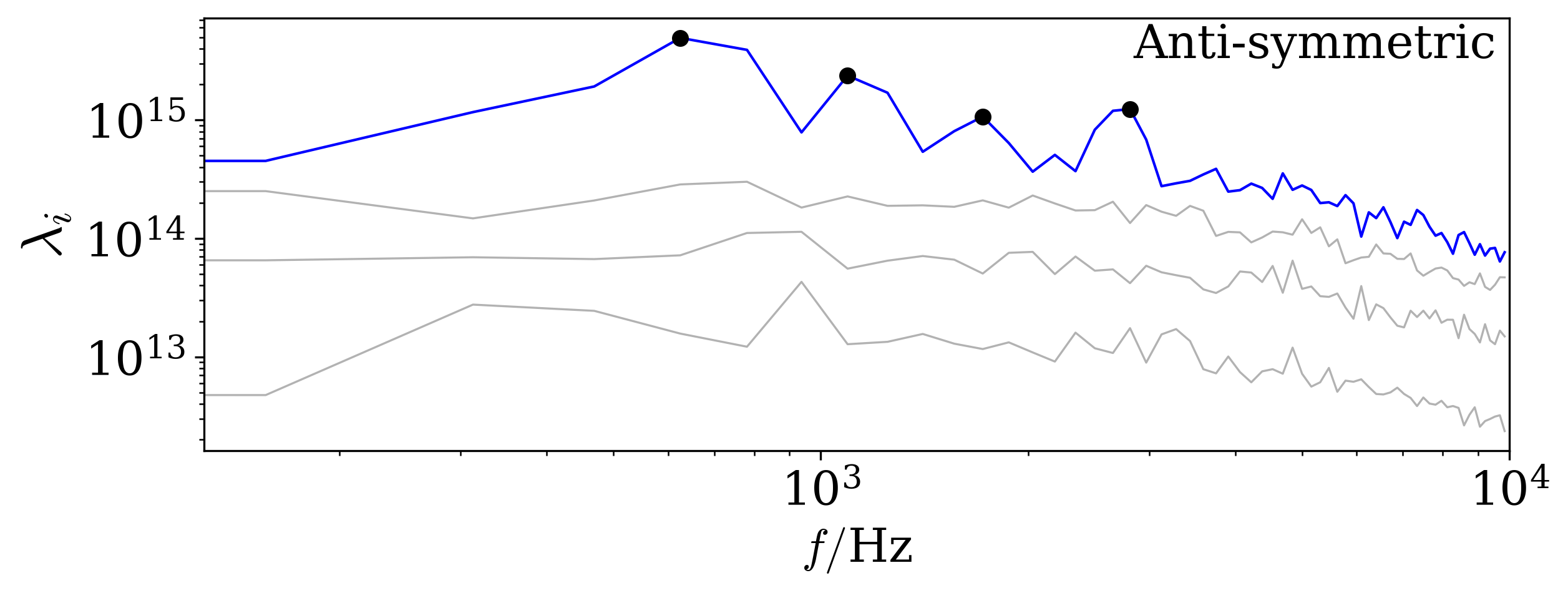}
  \label{fig:SPOD_sv_hr_as}
\end{subfigure}
\end{subfigure}

\caption{\footnotesize \gls{SPOD} eigenvalues of heat release fluctuations.}
\vspace{-0.5cm}
\label{fig:SPOD_hr_subfigs}
\end{figure}
\begin{figure*}[hbt]
\includegraphics[width=\textwidth]{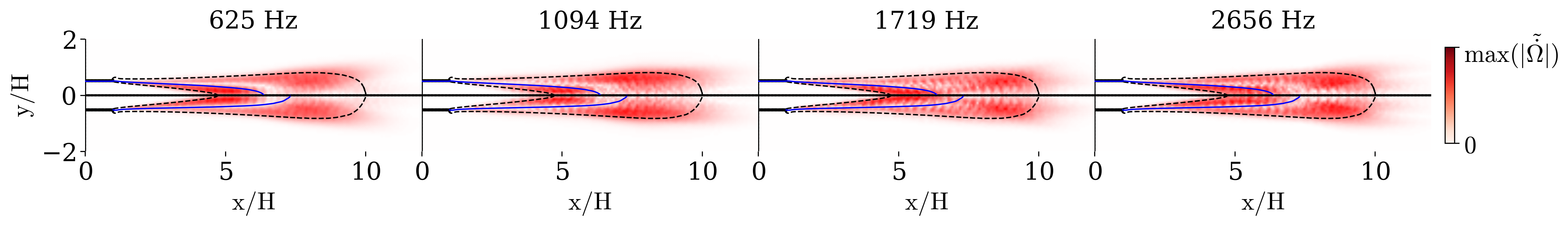}
\caption{\footnotesize \gls{SPOD} heat release distributions; turbulent flame brush: dashed black lines. Two neutral lines ($\widehat{\dot{\Omega}} = 0$): solid blue color. Top half: \gls{EBU} model neutral line ($\overline{c} \approx 0.36$); bottom half: algebraic model neutral line ($\overline{c} \approx 0.57$). }
\label{fig:spod_hr_mag_modes}
\end{figure*}

The \gls{SPOD} procedure described in Section~\ref{ch:theory:spod} has been applied to $300$ snapshots with a time increment of $\Delta t = 5\times10^{-5}\,\mathrm{s}$. Due to the comparably small number of available snapshots, the number of segments is limited to $N_\text{seg}=5$ with $66\%$ overlap, which maintains an adequately high frequency resolution. The investigated fields are velocity components $\widehat{u}_x$ and $\widehat{u}_y$, heat release rate $\widehat{h}$, and progress variable $\widehat{c}$. The dashed curves in Fig.~\ref{fig:gains_velocity} display the eigenvalues based on $u_x$ and $u_y$ for the symmetric and anti-symmetric cases, respectively. Both symmetries exhibit their highest spectral energy in the frequency interval $300\,\mathrm{Hz}$--$1000\,\mathrm{Hz}$. Above $1\,\mathrm{kHz}$, the spectral energy decays monotonically, as expected.

To investigate the turbulent flame dynamics, an analogous \gls{SPOD} analysis is performed on the fluctuations of the heat release rate. The corresponding eigenvalue spectra are displayed in Fig.~\ref{fig:SPOD_hr_subfigs}. The symmetric heat release fluctuations do not exhibit low-rank behavior. For anti-symmetric dynamics, the leading eigenvalue is larger than the symmetric counterpart and significantly dominates the sub-leading anti-symmetric eigenvalues. Thus, the anti-symmetric structures dominate the local heat release fluctuations, while these cannot contribute to global heat release fluctuations. Additionally, distinct peaks are observed at $625\,\mathrm{Hz}$, $1094\,\mathrm{Hz}$, $1719\,\mathrm{Hz}$, and $2656\,\mathrm{Hz}$. From here on, only the anti-symmetric fluctuations are considered, as they dominate the heat release response. The corresponding mode shapes will be discussed and compared to their \gls{RA} counterparts in Section~\ref{subsec: RAResults}.

\subsection{A Priori Analysis of the Flame Models \label{subsec: apriori analysis}} \addvspace{5pt}
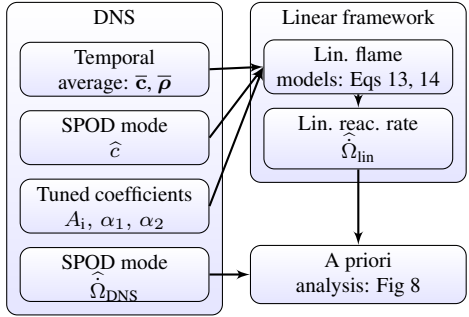
\begin{figure}[hbt]
    \centering
    \addvspace{-1.5cm}
    {\pgfplotsset{
    compat=1.3, 
    every axis/.append style={scale only axis,
    height=2.0cm, width=0.3\textwidth, xmin=0, xmax=2,ymin=0,ymax=1.2
    }
}
\footnotesize

\begin{tikzpicture}[node/.style={draw}]
\node (center) {};
 \node [rounded corners, top color=white, bottom color=blue!10,draw,draw,rectangle, minimum height=0.29\textwidth, minimum width=0.2\textwidth, anchor=north east] (LES) {};
 \node [anchor=north, at=(LES.north)] (LEStext) {DNS};
 \node [rounded corners, top color=white, bottom color=blue!10,draw,draw,rectangle, minimum height=0.167\textwidth, minimum width=0.2\textwidth, anchor=north west] (linFram) at ($(LES.north east)+(0.025\textwidth,0.00\textwidth)$) {};
 \node [anchor=north, at=(linFram.north), align = center] (linFramText) {Linear framework};

 \node [rounded corners, top color=white, bottom color=blue!10,draw,draw,rectangle, minimum height=0.05\textwidth, minimum width=0.175\textwidth, anchor=south] (FourierModes) at ($(LES.south)+(0.0\textwidth,0.0125\textwidth)$) {};
  \node [anchor=north, at=(FourierModes.north), align = center] (FourierModesText) {SPOD mode\\ $\widehat{\dot{\Omega}}_\text{DNS}$};
 \node [rounded corners, top color=white, bottom color=blue!10,draw,draw,rectangle, minimum height=0.05\textwidth, minimum width=0.175\textwidth, anchor=south] (tunedPrefactor) at ($(FourierModes.north)+(0.0\textwidth,0.0125\textwidth)$) {};
  \node [anchor=north, at=(tunedPrefactor.north), align = center] (tunedPefactorText) {Tuned coefficients\\ $A_\text{i}, \, \alpha_1, \,\alpha_2$};
  
   \node [rounded corners, top color=white, bottom color=blue!10,draw,draw,rectangle, minimum height=0.05\textwidth, minimum width=0.175\textwidth, anchor=south] (cfluc) at ($(tunedPrefactor.north)+(0.0\textwidth,0.0125\textwidth)$) {};
  \node [anchor=north, at=(cfluc.north), align = center] (cflucText) {SPOD mode\\ $
  \widehat{c}$};
 \node [rounded corners, top color=white, bottom color=blue!10,draw,draw,rectangle, minimum height=0.05\textwidth, minimum width=0.175\textwidth, anchor=south] (temporalMean) at ($(cfluc.north)+(0.0\textwidth,0.0125\textwidth)$) {};
  \node [anchor=north, at=(temporalMean.north), align = center] (temporalMeanText) {Temporal \\ average: $\overline{\mathbf{c}}$, $\overline{\bs{\rho}}$};

 \node [rounded corners, top color=white, bottom color=blue!10,draw,draw,rectangle, minimum height=0.05\textwidth, minimum width=0.175\textwidth, anchor=south, align = center] (linearResponse) at ($(linFram.south)+(0.0\textwidth,0.0125\textwidth)$) {Lin. reac. rate\\ $\widehat{\dot{\Omega}}_\text{lin}$};

  \node [rounded corners, top color=white, bottom color=blue!10,draw,draw,rectangle, minimum height=0.05\textwidth, minimum width=0.175\textwidth, anchor=south] (linEquation) at ($(linearResponse.north)+(0.0\textwidth,0.0125\textwidth)$) {};
  \node [anchor=north, at=(linEquation.north), align = center] (linEquationText) { Lin. flame \\ models: Eqs~\ref{eq:linEBU}, \ref{eq:linAlg} };
  
 \node [rounded corners, top color=white, bottom color=blue!10,draw,draw,rectangle,
 minimum width=0.2\textwidth, anchor= west, align = center] (apriori) at ($(FourierModes.east)+(0.0375\textwidth,0.0\textwidth)$) {A priori \\ analysis: Fig~\ref{fig:apriori_hr_modes}};
  

    \draw[->,>=latex',thick,draw=black,shorten >=0pt, shorten <=0pt] (temporalMean.east) to (linEquation.west);
    \draw[->,>=latex',thick,draw=black,shorten >=0pt, shorten <=0pt] (tunedPrefactor.east) to (linEquation.west);
    \draw[->,>=latex',thick,draw=black,shorten >=0pt, shorten <=0pt] (cfluc.east) to (linEquation.west);
    \draw[->,>=latex',thick,draw=black,shorten >=0pt, shorten <=0pt] (FourierModes.east) to (apriori.west);
    \draw[->,>=latex',thick,draw=black,shorten >=0pt, shorten <=0pt] (linEquation) to (linearResponse);
    \draw[->,>=latex',thick,draw=black,shorten >=0pt, shorten <=0pt] (linearResponse) to (apriori);
\end{tikzpicture} 

}
    \vspace{5pt}
    \caption{Schematic illustration of the a priori analysis.}
    \label{fig:apriori-schematic}
\end{figure}
The goal of the a priori analysis is to determine which of the reaction models described in Section~\ref{sec:BGandTH} is most consistent with the anti-symmetric \gls{SPOD} heat release modes. This approach allows testing a model in an isolated fashion without applying it in a closed set of equations, where secondary effects could complicate the model test. 

As a first step, we investigate the magnitudes of the \gls{SPOD} in anti-symmetric heat release fluctuations at four frequencies obtained from the \gls{DNS} as illustrated in Fig.~\ref{fig:spod_hr_mag_modes}. The heat release fields show regions of low activity within the turbulent flame brush, intersecting the flame brush in an upstream and a downstream region. This is in line with our expectations of the flame models under investigation in this study. As discussed in Section~\ref{sec:BGandTH} we expect the linearized flame models to show no flame activity where $\frac{\partial f}{\partial \overline{\Phi}} = 0$. For the \gls{EBU} model, this condition occurs at $\overline{c}_0 \approx 0.36$, whereas for the algebraic model, it occurs at $\overline{c}_{0} \approx 0.57$. To analyze which model is in better agreement with the \gls{SPOD}/\gls{DNS}, the heat release distributions in Fig.~\ref{fig:spod_hr_mag_modes} are overlaid with blue solid lines representing the iso-contours of $\overline{c} = \overline{c}_0$ for the \gls{EBU} model in the upper half plane and for the algebraic model in the lower half plane. In the remainder of this study, this region of zero flame activity will be referred to as the neutral line. The neutral line predicted by the algebraic model aligns with regions exhibiting negligible \gls{SPOD} heat release fluctuations, while the neutral line associated with the \gls{EBU} model intersects regions where significant \gls{SPOD} heat release fluctuations are present. This provides a first indication that the \gls{SPOD} heat release distributions align better with the algebraic model. 

Next, we progress to the a priori procedure as illustrated in Fig.~\ref{fig:apriori-schematic}. After tuning the model coefficients to the \gls{DNS}, the temporal mean fields and the \gls{SPOD} modes of the progress variable are introduced into the respective linearized reaction rate models. This yields the corresponding reaction rate fluctuations, denoted by $\hat{\dot{\Omega}}_\text{lin}$. This estimated quantity (A PRIORI-RR) is then compared with the \gls{SPOD} heat release modes.

\begin{figure*}[hbt]
\includegraphics[width=\textwidth]{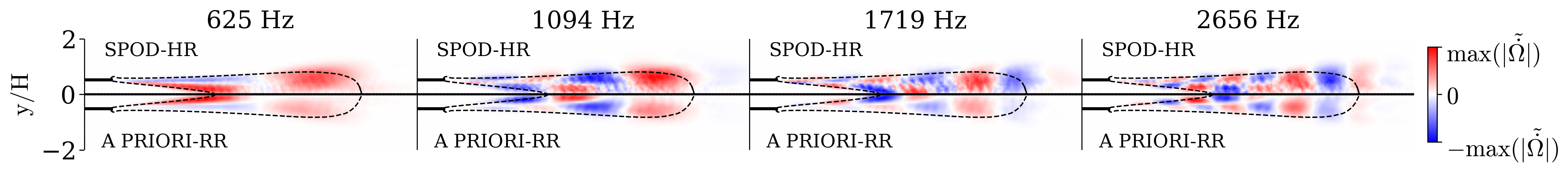}
\includegraphics[width=\textwidth]{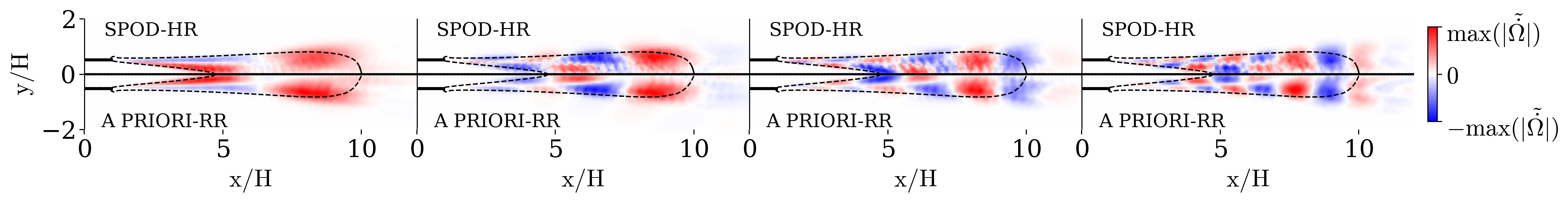}
\caption{\footnotesize Comparison of \gls{SPOD} heat release mode shapes (upper half planes) and a priori reaction rate modes (lower half planes), bordered by the turbulent flame brush (dashed black lines). Top row: \gls{EBU} model; Bottom row: algebraic model.}
\label{fig:apriori_hr_modes}
\end{figure*}

The outcomes of the a priori analysis are depicted in Figs~\ref{fig:apriori_hr_modes}, which shows that both models produce a priori reaction rate mode shapes that qualitatively resemble the corresponding \gls{SPOD} heat release modes. However, the modes predicted by the \gls{EBU} model decay further upstream compared with their \gls{SPOD} counterparts. In contrast, the mode shapes obtained from the algebraic model exhibit a decay rate that closely matches that observed in the \gls{SPOD} modes. Moreover, at the two higher frequencies, a slight misalignment appears near the symmetry axis between the a priori \gls{EBU} model modes and the corresponding \gls{SPOD} structures. For these same frequencies, the modes predicted by the algebraic model display better alignment with the \gls{SPOD} results.


\subsection{Resolvent Analysis results \label{subsec: RAResults}} \addvspace{10pt} 

\begin{figure*}[hbt]
\includegraphics[width=\textwidth]{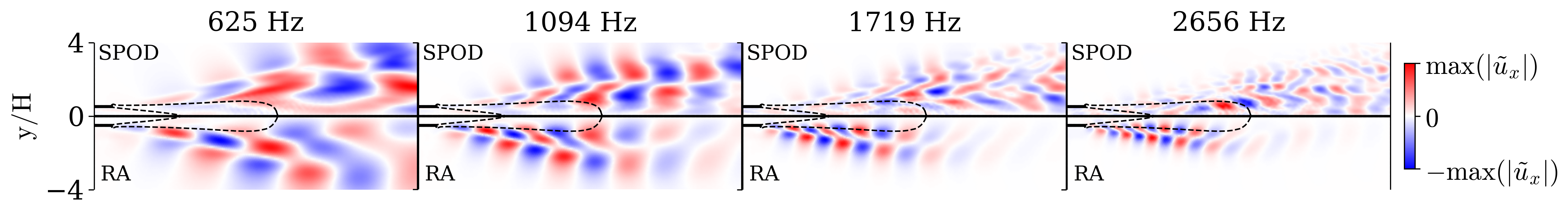}
\includegraphics[width=\textwidth]{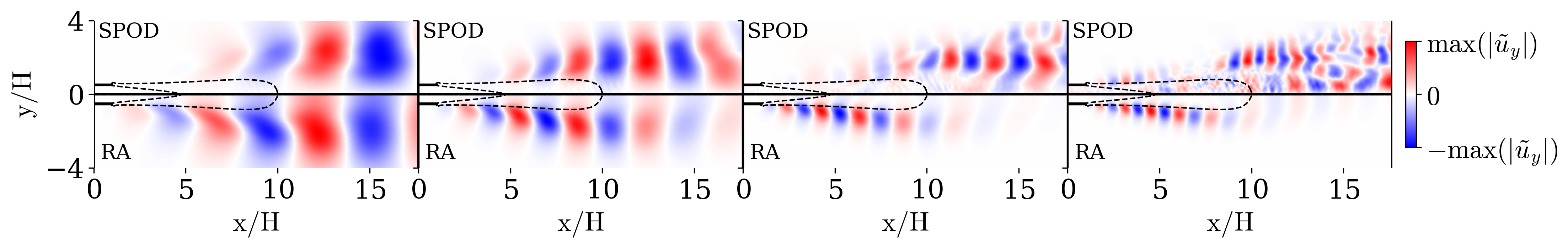}
\vspace{-4pt}
\includegraphics[width=\textwidth]{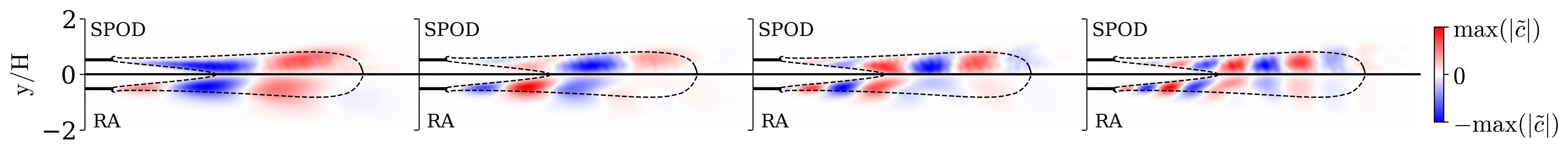}
\includegraphics[width=\textwidth]{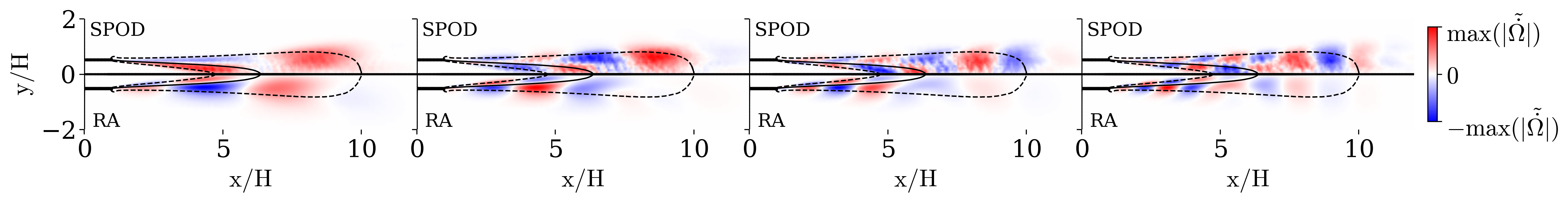}
\caption{\footnotesize Mode shapes from \gls{SPOD} (top half) and \gls{EBU} model-based \gls{RA} (bottom half): $\widetilde{u}_x$ (first row), $\widetilde{u}_y$ (second row), $\widetilde{c}$ (third row), and reaction rate (fourth row) overlaid with \gls{EBU} model neutral line ($\overline{c} \approx 0.36$) in solid black color.}
\label{fig:modes}
\end{figure*}
\begin{figure*}[hbt]
\includegraphics[width=\textwidth]{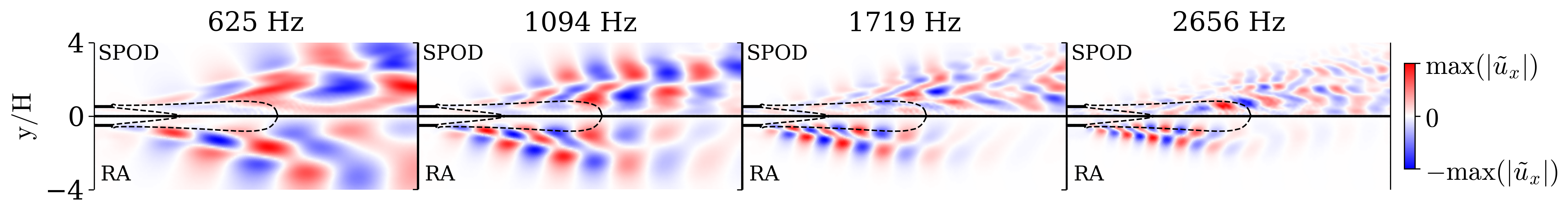}
\includegraphics[width=\textwidth]{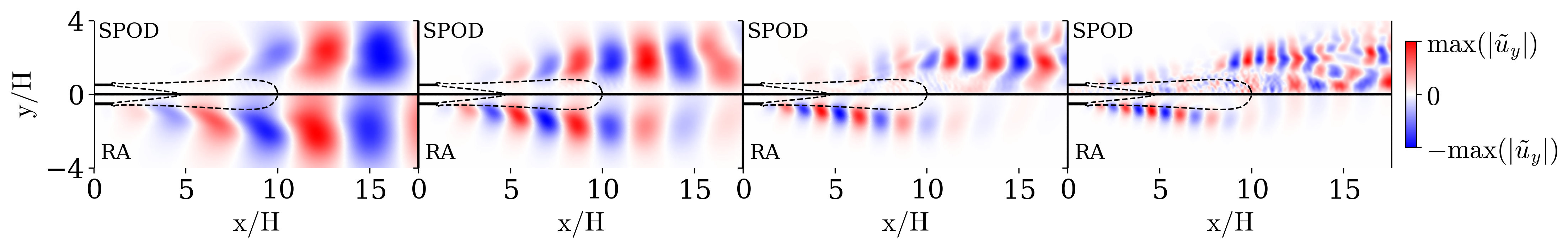}
\vspace{-4pt}
\includegraphics[width=\textwidth]{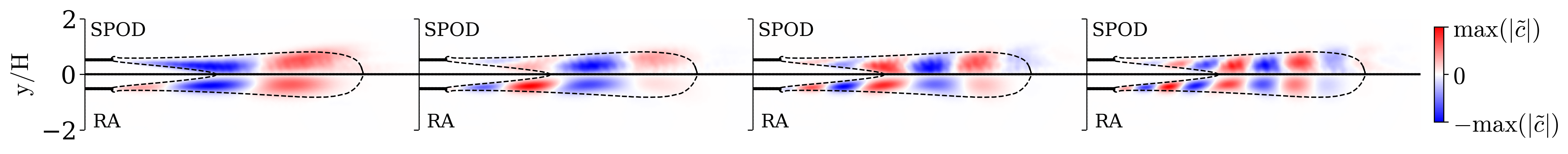}
\includegraphics[width=\textwidth]{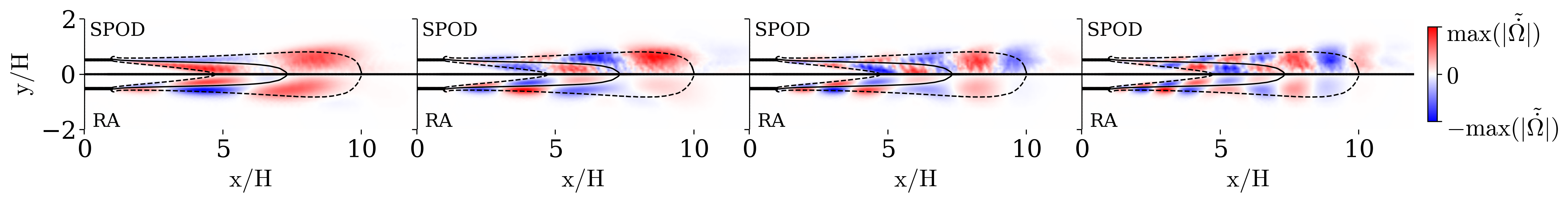}
\caption{\footnotesize Mode shapes from \gls{SPOD} (top half) and algebraic model-based \gls{RA} (bottom half): $\widetilde{u}_x$ (first row), $\widetilde{u}_y$ (second row), $\widetilde{c}$ (third row), and reaction rate (fourth row) overlaid with algebraic model neutral line ($\overline{c} \approx 0.57$) in solid black color.}
\label{fig:ahrm_modes}
\end{figure*}
\begin{figure*}[hbt]
\centering
\includegraphics[width=\textwidth]{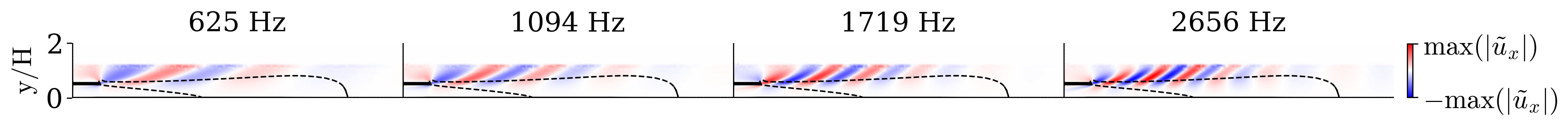}
\vspace{-4pt}
\includegraphics[width=\textwidth]{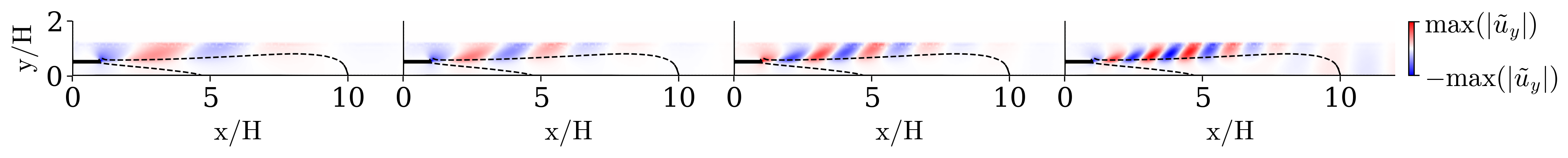}
\caption{\footnotesize Optimal forcing modes in the streamwise velocity (first row) and cross streamwise velocity (second row).}
\label{fig:forcing_ebum}
\end{figure*}

\gls{RA} is performed for both symmetric and anti-symmetric configurations, where forcing is restricted to a box of length $12.5$H in x--direction, and height $1.25$H in y--direction. The resulting gains are shown as solid lines in Fig.~\ref{fig:gains_velocity}. The \gls{RA} spectra agree well with the \gls{SPOD} results and indicate low-rank behavior for both configurations. The strongest amplification occurs in the frequency range $300\text{–}1000~\text{Hz}$, consistent with the \gls{SPOD} spectra.


The anti-symmetric \gls{RA} modes of the velocity components, progress variable, and reaction rate (top to bottom) are compared against their \gls{SPOD} counterparts in Fig.~\ref{fig:modes} for the linearized \gls{EBU} model. The results are shown for the four frequencies, which correspond to the peaks in the \gls{SPOD} of the heat release spectrum illustrated in Fig.~\ref{fig:SPOD_hr_subfigs}, which correspond to the same frequencies illustrated in Fig.~\ref{fig:spod_hr_mag_modes}. The upper half planes in Fig.~\ref{fig:modes} illustrate the SPOD modes, while the lower half planes show the corresponding \gls{RA} mode shapes. The equivalent illustration for the algebraic model is shown in Fig.~\ref{fig:ahrm_modes}. The \gls{SPOD} velocity modes exhibit the characteristic Kelvin-Helmholtz roll-up pattern \cite{schmidt2018spectral} at the two lower frequencies \cite{ZHENG2018683}. For the higher frequencies, however, the \gls{SPOD} analysis does not clearly recover this pattern. It is likely that this is due to the very short data set available for the analysis in combination with the significantly lesser energy contained in this frequency range, as can be seen in the \gls{SPOD} spectrum in Fig.~\ref{fig:gains_velocity}. For the two lower frequencies, the \gls{RA} velocity mode shapes agree well with the \gls{SPOD} modes. At higher frequencies, the decay of the \gls{SPOD} modes makes direct comparison difficult. Overall, the velocity modes appear largely insensitive to the choice of reaction model.

We further compare \glspl{RA} using the two flame models for fluctuations in the progress variable and heat release. The \gls{EBU} model qualitatively reproduces the \gls{SPOD} mode shapes. However, especially at higher frequencies, the algebraic model shows significantly better agreement with the reference results, consistent with the finding of the a priori analysis. In particular, the discrepancies near the neutral line, where no reaction rate is expected, are notably reduced. These results indicate that the high-fidelity physics contained in the \gls{DNS} mean state can be partly captured by fitting an algebraic model to the data, such that its linearization provides reasonably accurate predictions of the reaction dynamics.

Finally, Figs.~\ref{fig:forcing_ebum} show the optimal \gls{RA} forcing modes, restricted to a box of $12.5\text{H}\times1.25\text{H}$ for the four previously discussed frequencies. The first row displays the streamwise velocity component, and the second row the cross-streamwise component. The results indicate that the dominant flow structures are most receptive to turbulent forcing in the free-stream region outside the turbulent flame brush, which differs from the observations reported in earlier studies. For instance, Lesshafft et al.~\cite{PhysRevFluids.4.063901} found that optimal forcing in circular non-reacting jets occurs in the nozzle shear layer, while Casel et al.~\cite{casel2022resolvent} showed that in the circular reacting jet configuration they investigated, it is concentrated near the coaxial pilot burner.

\section{Conclusion \label{conclusion}} \addvspace{5pt}

This study applies two active-flame resolvent frameworks to a turbulent hydrogen–air slot flame: a statistics-based algebraic model that is calibrated using high-fidelity \gls{DNS} data and a physics-based \gls{RANS}-\gls{EBU} model. The resulting linear input–output predictions are compared with \gls{SPOD} modes extracted from the same \gls{DNS} dataset, where anti-symmetric dynamics are identified as the dominant contributor to heat release fluctuations. Notably, within these anti-symmetric dynamics, the \gls{SPOD} spectra associated with heat release exhibit a pronounced tonal behavior, whereas the corresponding velocity eigenvalues display a broadband response. First, it was demonstrated with an a priori analysis that the algebraic model outperforms the \gls{EBU} model in reproducing the high-fidelity data. Then, an a posteriori test was performed for both models. Here, for both models the gain spectra and velocity mode shapes show good agreement between \gls{RA} and \gls{SPOD} within the main energetic frequency range of $300–1000 \text{ Hz}$. At higher frequencies, however, the alignment of progress variable and reaction rate modes with their \gls{SPOD} counterparts improves noticeably when the algebraic model is used in the linearized \gls{RA} framework. These results indicate that a nonlinear algebraic model calibrated with high-fidelity \gls{DNS} data remains effective and relevant in the linearized framework and improves the predictive capability of active-flame \gls{RA}, potentially in different turbulent reacting configurations.

\acknowledgement{CRediT authorship contribution statement} \addvspace{5pt}
{\bf Anant Rajeev Talasikar}: Investigation, Software (\gls{FELiCS}), Visualization and Formal Analysis. {\bf Marina Matthaiou}: Validation and Formal analysis. {\bf Michael Gauding}: Resources and Software (DNS). {\bf Heinz Pitsch}: Resources and Funding acquisition. {\bf Thomas Ludwig Kaiser}: Conceptualization, Supervision and Funding acquisition.

\acknowledgement{Declaration of competing interest} \addvspace{5pt}


The authors declare that they have no known competing financial interests or personal relationships that could have appeared to influence the work reported in this paper.

\acknowledgement{Acknowledgments} \addvspace{5pt}
The authors acknowledge support from the German Federal Ministry of Research, Technology and Space and the federal states through the National High-Performance Computing (NHR) program on the CLIAX supercomputer at RWTH Aachen University. 

Funded by the Deutsche Forschungsgemeinschaft (DFG, German Research Foundation) – 530442286, 523874889, 523881008.


\footnotesize
\baselineskip 9pt

\clearpage
\thispagestyle{empty}
\bibliographystyle{proci}
\bibliography{PROCI_LaTeX}


\newpage

\small
\baselineskip 10pt


\end{document}